%%%%%%%%%%%%%%%%%%%%%%%%%%%%%%%%%%%%%%%%%%%%%%%%%%%%%%%%%%%%
%                                                         
%
%             This file  uses the macro package            %
%                                                          % 
%                      PHYZZX.TEX                          %  
%                                                          %  
%%%%%%%%%%%%%%%%%%%%%%%%%%%%%%%%%%%%%%%%%%%%%%%%%%%%%%%%%%%%

\def\CC{{\mathchoice
{\rm C\mkern-8mu\vrule height1.45ex depth-.05ex 
width.05em\mkern9mu\kern-.05em}
{\rm C\mkern-8mu\vrule height1.45ex depth-.05ex 
width.05em\mkern9mu\kern-.05em}
{\rm C\mkern-8mu\vrule height1ex depth-.07ex 
width.035em\mkern9mu\kern-.035em}
{\rm C\mkern-8mu\vrule height.65ex depth-.1ex 
width.025em\mkern8mu\kern-.025em}}}

\input phyzzx.tex
%\draft

\def\np{Nucl. Phys.}
\def\pl{Phys. Lett.}

\def\pr{Phys. Rev.}

\tolerance=500000
\overfullrule=0pt
\pubnum={US-FT-29/97\cr hep-th/9709191}
\date={September, 1997}
\pubtype={}
\titlepage

\title{ON T-DUALITY IN DILATONIC GRAVITY} 
\author{Pablo M. Llatas\foot{E-mail: llatas@fpaxp1.usc.es}}
\address{Departamento de F\'\i sica de
Part\'\i culas, \break Universidad de Santiago, \break
E-15706 Santiago de Compostela, Spain. \break}

\abstract{Under the assumption of axial symmetry we introduce a map 
from dilatonic gravity to a string-like action. This map allows
one to introduce, in a rather simple way, the equivalent 
of string theory T-duality in dilatonic gravity. Here we choose
the duality group to be an $SO(2,1)$ group and, for a particular 
rotation, we recover a symmetry of dilatonic gravity discussed previously
in the literature.}

\endpage
\pagenumber=1
\sequentialequations

%\sequentialequations

Recently, there has been a serious improvement in the understanding
of black-hole physics in the context of string theory. Near the 
singularity inside a black-hole, the classical theory of gravity breaks
down and one expects quantum gravity processes to take place. Here is
where string theory comes to play a role. In particular, the origin of
Bekenstein-Hawking entropy for certain class of black-holes has
been successfully explained by a counting of quantum microstates
in the framework of string theory
\REF\Stro{A. Strominger and C. Vafa \journal\pl&B379(96)99.}
\REF\Maldacena{J.M. Maldacena, "Black-holes in String Theory", 
Ph. D. Thesis, Princeton University,
 (hep-th/9607235) and references therein.}
[\Stro ,\Maldacena]. Those special black-holes, in the
simplest models, are 
typically the ones with Ramond-Ramond fields excited (in order to
stabilize the dilaton on the horizon) and with a BPS inequality
saturation condition (in such a way that masses and charges are
protected from quantum corrections and so one can identify string
perturbative states with the black-hole solutions of the 
effective theory).

Despite of the fact that the formalism of the counting sits over 
the string theory
framework, it can be extrapolated to 
some of the most popular black-hole solutions of General Relativity. 
For instance, the standard five-dimensional extremal Reissner-Nordstrom
metric can be understood as an 
specific combination of a number of D-fivebranes
and D-onebranes with momenta in a compactified dimension 
[\Maldacena].  For this specific
combination, the dilaton field turns out to be a constant, 
and one recovers the
Einstein-Maxwell theory. Due to this success of string theory, 
one can ask
which properties of string theory, 
specially duality, can be translated (in the closest analogy with 
string theory) to
non-string (or apparently  non-string) situations.

The importance of duality is two-fold. First, 
it relates classical solutions of a given theory 
through duality transformations. 
For instance, the dilaton 
Melvin solutions of Gibbons-Maeda 
\REF\Gibbons{G.W. Gibbons and K. Maeda \journal\np&B298(88)741.}
[\Gibbons] is
related through T-duality to the trivial flat space solution, as
we will see bellow (see also \REF\Quevedo{F. Quevedo, "Duality
and Global Symmetries", hep-th/9706210.}
\REF\Dowker{F. Dowker, J.P. Gauntlett,
D.A. Kastor and J. Traschen \journal\pr&D49(94)2909.}
[\Quevedo ,\Dowker]). 
Second, one can use duality to generate
new solutions of the equations of motion corresponding 
to a given action. 
 
In this paper we will study T-duality 
\REF\Yoli{E. Alvarez, L. Alvarez-Gaume and Y. Lozano
\journal\np&{\sl Proc. Suppl.} 41(95)1 and references therein.}[\Yoli]
in the theory described 
by the action:

$$
S=\int{d^4 x \sqrt{-G} \Bigl ( R^{G} -{1\over 2} G^{\mu\nu}
\partial_{\mu}\Phi\partial_{\nu}\Phi -{1\over 4} e^{-a\Phi}
F^2_G \Bigr ), }
\eqn\uno
$$
where, because we are going to define different metrics through the
paper, we have
explicitly written as subindices and superindices the metric
employed to construct the corresponding quantities.

For $a=0$ this is the usual Einstein-Maxwell action. For $a=1$  
is part of the low energy action of string theory and for 
$a=\sqrt{3}$ is the dimensional reduction of five-dimensional 
Kaluza-Klein theory. We will work with the case of arbitrary $a$ 
\REF\Garflinke{D. Garfinkle, G. Horowitz and A. Strominger 
\journal\pr&D43(91)3140 and {\bf 45}(1992)3888(E).}
([\Garflinke]). We would like to study the analog of string T-duality  
in this (non-string) system.
To simplify the problem, we are going to restrict ourselves 
to the classical solutions of the equations of motion of the
previous action which are independent of an angular coordinate
which we call $\varphi$ and such that $G_{i\varphi}=A_{i}=0$
(from now on, $i=0,1,2$ denotes the other three coordinates). In fact,
all the program can be carried out by selecting any other 
isometry of the metric (or more than one isometry at the same time). 
Moreover, one could relax
the previous conditions  on  $G_{i\varphi}$ and $A_{i}$.
Here, for simplicity, we will work
only this particular case  and more general situations will 
be considered elsewhere.

As a first step, we perform a local conformal transformation on
the metric $G_{\mu\nu}$ by using the dilaton field, 
and define $g_{\mu\nu}$ by:

$$
G_{\mu\nu}=e^{\alpha\Phi} g_{\mu\nu}
\eqn\dos
$$

In terms of $g_{\mu\nu}$ the action \uno\ reads:

$$
S=\int{d^4 x \sqrt{-g} e^{\alpha\Phi}(R^{g} +{1\over 2} 
(3{\alpha}^2-1)g^{\mu\nu}
\partial_{\mu}\Phi\partial_{\nu}\Phi -{1\over 4} e^{-(a+\alpha)\Phi}
F^2_g)}
\eqn\tres
$$

As the solutions we are about to work with are independent
of the coordinate $\varphi$, we can dimensionally reduce the theory
from four to three dimensions by integrating $\varphi$. The
three dimensional action then reads:

$$
\eqalign{
S=&{2\pi\over B}\int d^3 x \sqrt{-g} 
e^{(\alpha\Phi +{1\over 2} \ln{(B^2
g_{\varphi\varphi})})} (R^{g} +\alpha g^{ij}\partial_i \Phi{\partial_j 
g_{\varphi\varphi}\over g_{\varphi\varphi}}+\cr
&+{1\over 2} (3{\alpha}^2-1) g^{ij}
\partial_{i}\Phi\partial_{j}\Phi -{1\over {2g_{\varphi\varphi}}}
e^{-(a+\alpha)\Phi} 
g^{ij}\partial_i A_{\varphi}\partial_j A_{\varphi})\cr
}
\eqn\cuatro
$$

The $B$ parameter has the dimension of $\sqrt{g^{\varphi\varphi}}$ 
(i.e., the dimension of a magnetic field) and 
has been introduced to put an adimensional 
quantity inside the logarithm. 
Every quantity in the previous action (such as the 
Riemann scalar curvature and
the volume element) refers now to the three dimensional metric defined 
by $g_{ij}$.

Let us now introduce the fields $w$, $v$ and $\bar\Phi$ defined by:

$$
\eqalign{
w&\equiv A_{\varphi},\cr
v&\equiv 4 g_{\varphi\varphi} e^{(a+\alpha )\Phi},\cr
\bar\Phi &\equiv {2\over\sqrt{1+a^2}} (\alpha\Phi +{1\over 2}
\ln{B^2 g_{\varphi\varphi}}).\cr }
\eqn\cinco
$$

If we express the action \cuatro\ 
in terms of these fields, one can check 
that if and only if we fix $\alpha=-{1\over a}$,
we can reduce the coupling of $\bar\Phi$ to all other fields 
to a global exponential factor. For this value of $\alpha$ 
(called the ``total metric" in [\Dowker]) the action reads:

$$
\eqalign{
S &\propto \int d^3 x \sqrt{-g}
e^{{\sqrt{1+a^2}\over 2} \bar\Phi}(R^g +{1\over 2} 
g^{ij}\partial_{i}{\bar\Phi}\partial_{j}{\bar\Phi}-{1\over {2(1+a^2)}}
g^{ij}{\partial_{i} v\over v}{\partial_{j}v\over v}-{2\over v}
g^{ij}\partial_{i} w\partial_{j} w)\cr
  &= \int d^3 x \sqrt{-g}
e^{{\sqrt{1+a^2}\over 2} \bar\Phi} \bigr ( R^g +{1\over 2} 
g^{ij}\partial_{i}{\bar\Phi}\partial_{j}{\bar\Phi}+{1\over {2(1+a^2)}}
g^{ij}\partial_i ({1\over v})\partial_j \Bigl (v+2(1+a^2) w^2
\Bigr )\cr
&\qquad\qquad -2 g^{ij}\partial_i 
({w\over v})\partial_j w \bigl ). \cr
}
\eqn\seis
$$

Note that in the last equality, we have written the action as a
sum of products of derivatives. This is so because we want to
write the two last terms as the trace of a matrix $M$:

$$
{1\over 8} g^{ij}Tr(\partial_i M\partial_j M)={1\over {2(1+a^2)}}
g^{ij}\partial_i ({1\over v})\partial_j \Bigl ( v+2(1+a^2) w^2\Bigr )
-2 g^{ij}\partial_i ({w\over v})\partial_j w .
\eqn\siete
$$

If we find such a matrix $M$, the action would take a similar 
form to the bosonic part of the low energy 
effective action of the string. 
Actually, we would have:

$$
S=\int d^3 x\sqrt{-g} e^{{\sqrt{1+a^2}\over 2}\bar\Phi}\bigl ( R^g+
{1\over 2}g^{ij}\partial_{i}{\bar\Phi}\partial_{j}{\bar\Phi}+
{1\over 8} g^{ij}Tr(\partial_i M\partial_j M)\Bigr ).
\eqn\ocho
$$

This action has an obvious symmetry under the transformations:

$$
\eqalign{
g_{ij} &\longrightarrow g_{ij}'=g_{ij} ,\cr
\bar\Phi &\longrightarrow \bar\Phi '=\bar\Phi ,\cr
M &\longrightarrow M'=\Omega M\Omega^{-1}, \cr
}
\eqn\nueve
$$
being $\Omega$ a global transformation. This transformation is 
the analog
of the T-duality transformation of the string. The difficult point is 
to find a matrix $M$ fulfilling equation \siete\ and covariant under 
the previous symmetry. What we mean in this last 
statement is the following: 
$M$ should have entries which are functions of $v$ and $w$ 
(and so, through
\cinco, functions of the scalars of the theory $\Phi$, 
$A_{\varphi}$ and
$g_{\varphi\varphi}$). By covariance we mean that the entries of the
rotated matrix $M'$ are the $same$ functions, but now 
with the scalars $v$ and $w$ substituted by  
$v'$ and $w'$ respectively (which represent the dual 
fields with respect to
$v$ and $w$). If the duality group $\cal G$ to which 
$\Omega$ belongs is
generated by a set of generators $X_i$, the covariance can 
be summarized
in the following relation:

$$
[M,X_i]={\partial M\over \partial v}\delta_i v+
{\partial M\over \partial w}\delta_i w\qquad\forall i,
\eqn\diez
$$
where $\delta_i v$ and $\delta_i w$ are the infinitesimal variations
of the fundamental fields under the action of 
the generator $X_i$. Given 
the generators $X_i$, the equation \diez\ is a system of first order
differential equations for the entries of $M$ that can be written 
straightforwardly. Also, the part of the off-diagonal elements of
$M$ can be inferred from the equation \siete.

Let us make a particular election of the group $\cal G$. 
First, from the
structure of equation \siete, we see that the minimum 
size of the matrix 
$M$ is three
(in order to accommodate the four different derivatives appearing 
in equation \siete\
 in four  off-diagonal entries of $M$). Thus, let us take $M$ 
as a three 
dimensional matrix (therefore, $\Omega$ is a three dimensional 
representation 
of the group $\cal G$). The second observation is that the duality 
relation
\nueve\ does not depend on the determinant of $\Omega$. So the group 
$\cal G$
can be chosen as a subgroup of $SL(3,R)$. Here, 
inspired in the string case, we will choose $SO(2,1)$ with generators
(in the adjoint representation):

$$
X_1 =\pmatrix{0 & 0 & -1\cr
             1 & 0 & 0\cr
             0 & 0 & 0\cr};\,\,
       X_2 =\pmatrix{0 & -1 & 0\cr
             0 & 0 & 0\cr
             1 & 0 & 0\cr};\,\,
         \,\, X_3 =\pmatrix{0 & 0 & 0\cr
             0 & 1 & 0\cr
             0 & 0 & -1\cr}.
\eqn\once
$$
In this case, equations \siete\ and \diez\ can be 
easily solved. The solution for the
matrix $M$  is given by:

$$
\matrix{
m_{11}=c+{4\over c} {w^2\over v}; \qquad\hfill 
& m_{12}=2 B \sqrt{c} (w+{2\over c^2}{w^3\over v});\hfill
& m_{13}=-{2\over {B\sqrt{c}}} {w\over v}; \hfill\cr 
m_{21}=m_{13};\hfill 
& m_{22}=-{2\over c}{w^2\over v};\hfill 
& m_{23}={1\over {B^2 v}};\hfill\cr
m_{31}=m_{12};\hfill 
& m_{32}=4 B^2 ({c^2\over 4}v+w^2+
{1\over c^2}{w^4\over v});\qquad\hfill
& m_{33}=m_{22};\hfill\cr
}
\eqn\doce
$$
being $c$ a number depending on the parameter $a$:

$$
c\equiv\sqrt{{2\over {1+a^2}}}.
\eqn\trece
$$
If we would have chosen another basis for the $SO(2,1)$ generators,
we would have obtained the corresponding $SO(2,1)$-rotation
of the matrix $M$ as a solution of \diez .
Note that we have conveniently introduced the parameter $B$ 
(with dimensions of a magnetic field) in the entries
of $M$ to make them adimensional, in such a way that the duality 
transformation \nueve\
is well defined. One can check that the matrix $M$ previously defined
satisfies \siete\ and \diez. In fact, for $a=1$ (the string case),
the entries \doce\ are the axial-symmetric direct analog
of the definitions used by Sen in 
\REF\Sen{A. Sen \journal\np&B440(95)421.} [\Sen] 
(see also \REF\Schwarz{J. Maharana and J. Schwarz \journal\np&
B390(93)3.} [\Schwarz]) to study
T-duality in the heterotic string compactified on 
a torus. Note that the
isometry used there by Sen was the time-independence of the solutions 
and we are
using  the $\varphi$-independence and, as we said before, the
formalism used here can be generalized to any isometry.

Now, we are ready to perform a duality transformation of 
the type \nueve . We
just have to use any element of the group $SO(2,1)$ generated by $X_1$,
$X_2$ and $X_3$. The result will be a matrix $M'$ whose entries 
$m'_{23}$ and $m'_{13}$ are related to the dual fields $v'$ and $w'$ 
by (see equation \doce ):

$$
v'={1\over {B^2 m'_{23}}};\qquad\qquad 
w'=-{\sqrt{c}\over 2B}{m'_{13}\over
{m'_{23}}}. \eqn\catorce
$$
Now, one can go backwards in our formulae 
and express the dual fields in
terms of the fields appearing in the first action we wrote in 
equation \uno. 
After some
algebra, one obtains:

$$
\eqalign{
\Phi ' &=\Phi -{2a\over {1+a^2}}\ln\Lambda,\cr
A_{\varphi}' &=-{\sqrt{c}\over {2B}} {m_{13}'\over m_{23}'},\cr
G_{\varphi\varphi}' &=\Lambda^{-{2\over {1+a^2}}} 
G_{\varphi\varphi},\cr
G_{ij}' &=\Lambda^{2\over {1+a^2}} G_{ij},\cr
}
\eqn\quince
$$
where $\Lambda$ is given by:

$$
\Lambda =2 B\sqrt{m_{23}' G_{\varphi\varphi}}\, e^{{a\over 2}\Phi}.
\eqn\dieciseis
$$

These duality transformations have the same form that the ones 
introduced by Dowker et al. in [\Dowker]. But there, $\Lambda$ had the
following specific form ($\Lambda_D$):

$$
\Lambda_D ={\Bigl ( 1+{1+a^2\over 8} H A_{\varphi}\Bigr )}^2 +
{1+a^2\over 16} H^2 G_{\varphi\varphi} e^{a\Phi},
\eqn\dsiete
$$
(note that we are using different normalizations for the dilaton
and gauge field with respect to the ones used in [\Dowker]).
We want to relate this transformation with one of the $SO(2,1)$ 
transformations that we have just introduced. In order
to do this we first have to relate the parameter $H$ appearing in
$\Lambda_D$ with the parameters of our formalism ($B$ and the group
parameters of the $SO(2,1)$ transformation). Actually, if one
performs a $\Lambda_D$ rotation over the flat-space solution
in cylindrical coordinates, one obtains the dilaton Melvin
solutions of Gibbons-Maeda [\Gibbons]. This describes 
the closest analog
of a universe with a constant magnetic field in general relativity.
The $H$ parameter 
corresponds in this case 
to the strength of the magnetic field on the
axis ($H^2 ={1\over 2} F_{\mu\nu}F^{\mu\nu}\mid_{\rho=0}$).
We will perform an $SO(2,1)$ transformation generated by $X_1$
with parameter $\tau$. In this case, due to that $X_1^3=0$, 
it is very easy to write the matrix $\Omega$. The result is:

$$
\Omega_{\tau} =e^{\tau X_1}=\pmatrix{1 & 0 & -\tau\cr
                               \tau & 1 & -{\tau^2\over 2}\cr
                               0 & 0 & 1\cr}.
\eqn\docho
$$
Performing this $\Omega$ rotation in our matrix defined in \doce\ 
we obtain:

$$
\Lambda_{\tau}=2 B\sqrt{m_{23}'G_{\varphi\varphi}}e^{{a\over 2}\Phi}
=\Bigr ( 1+ {\tau B\over \sqrt{c}} A_\varphi\Bigl )^2
  + 2 B^2 \tau^2 c G_{\varphi\varphi} e^{a\Phi},
\eqn\dnueve
$$
and
$$
A^{(\tau)}_{\varphi}={1\over\Lambda_{\tau}} 
(A_{\varphi} +2 B\tau c^{3/2}
G_{\varphi\varphi} e^{a\Phi} +{B\tau\over \sqrt{c}}A_{\varphi}^2 ).
\eqn\veinte
$$

Putting $\Phi=A_{\varphi}=0$ and 
$G_{\varphi\varphi}=r^2$ in these equations, we obtain 
the dual solution of the
flat-space metric under $\Omega_{\tau}$. The result is:

$$
\eqalign{
\Lambda_{\tau} & =1+2 B^2 \tau^2 c r^2,\cr
A^{(\tau)}_{\varphi} & ={1\over\Lambda_{\tau}}2 B\tau c^{3/2} r^2.\cr}
\eqn\vuno
$$
$A^{(\tau)}_{\varphi}$ leads to a magnetic field on the axis:

$$
H =\sqrt{{1\over 2} F_{\mu\nu}F^{\mu\nu}}\mid_{\rho=0}=4 B\tau c^{3/2}.
\eqn\vdos
$$
This equation relates the parameter $B$ appearing in the definition
\doce\ of the matrix $M$ and the parameter 
$\tau$ of the gauge $SO(2,1)$
transformation with the magnetic field of the dilaton Melvin 
solution on the axis ($r=0$). Writing now the relations \dnueve\ in
terms of $H$, we see that the $\tau$ parameter disappears from
the equations and we get:

$$
\eqalign{
\Lambda_{\tau} &={\Bigl ( 1+{1+a^2\over 8} H A_{\varphi}\Bigr )}^2 +
{1+a^2\over 16} H^2 G_{\varphi\varphi} e^{a\Phi}=\Lambda_D,\cr
A^{(\tau)}_{\varphi} & ={1\over\Lambda_{\tau}}
(A_{\varphi} +{H\over 2}
G_{\varphi\varphi} e^{a\Phi} +{H\over {4c^2}}A_{\varphi}^2 ).\cr}
\eqn\vtres
$$
Finally, performing a gauge transformation 
$A_{\mu}\rightarrow A_{\mu}+\partial_{\mu}\lambda$ with 
$\lambda$ being:
$$
\lambda= -{4 c^2 \over H}\varphi,
\eqn\vcuatro
$$
we recover the transformations \dnueve\ and \veinte\ as in [\Dowker].
We have then identified the uniparametric subgroup of $SO(2,1)$ 
generating these transformations.

As we mentioned above, the same program we have carried out 
using the isometry generated by $\varphi$-translations can
be performed by using the isometry under time-translations. If
we do so, instead of magnetic charges we will obtain electric 
charges. This, in turn, is equivalent to perform an electric-magnetic
duality in the transformations described above. In fact, the action
\uno\ is invariant under this electric-magnetic duality,
which is given by:

$$
{\hat F}_{\mu\nu}={1\over 2} e^{-a\Phi}\epsilon_{\mu\nu\rho\sigma}
F^{\rho\sigma},\qquad\qquad {\hat\Phi}=-\Phi.
\eqn\vcinco
$$
\vskip2cm

We have studied T-duality in dilatonic gravity and found a
$SO(2,1)$ invariance of the theory. A previously 
discussed transformation
in the literature has been recovered as a particular
$SO(2,1)$ rotation in our formalism.
There are several directions in which this formalism can be 
generalized. First, one could perform a second dimensional
reduction (by integrating the time coordinate) to obtain 
a formalism collecting T-duality and electric-magnetic
duality together and on the same footing. 
In this case the
matrix $M$ will be, at least, five-dimensional. 
This could be useful to study
dyon solutions of dilatonic gravity. In addition, one could
relax the condition $G_{t\varphi}=0$. This would provide us
with tools to implement duality transformations 
on solutions with rotation. 
Also, one could try to solve
the equations \siete\ and \nueve\ for other group different 
from $SO(2,1)$ giving new dualities
in dilatonic gravity. Finally, one could try to extend this
formalism for non-abelian gauge groups. In this case, the presence
of non-derivative terms on the gauge field in the action will
imply that powers of the matrix $M$ will also appear  in the 
action. We expect to report on these topics elsewhere.

\ack

I am indebted with I.P Ennes, J. S\'a nchez Guill\' en
J. S\' anchez de Santos and A.V. Ramallo  
for useful comments.
This work has been partially supported by CICYT AEN96-1673
and European Comunity TMR ERBFMRXCT960012.

\refout

\end